\documentstyle[aps]{revtex}

\newcommand{\beq}{\begin{equation}}
\newcommand{\eeq}{\end{equation}}
\newcommand{\bra}[1]{\left< {#1} \right|}
\newcommand{\ket}[1]{\left| {#1} \right>}
\newcommand{\tg}{\tilde{g}}

\begin{document}
\draft

\title{Simple experimental methods for trapped ion quantum processors}

\author{D. Stevens, J. Brochard and A. M. Steane}

\address{Department of Physics, Clarendon Laboratory,
Parks Road, Oxford, OX1 3PU, England}

\date{January 1998}

\maketitle

\begin{abstract}
Two techniques are described that simplify the experimental requirements
for measuring and manipulating quantum information stored in trapped ions.
The first is a new technique using electron shelving to measure the
populations of the Zeeman sublevels of the ground state, in an ion for
which no cycling transition exists from any of these sublevels. The second
technique is laser cooling to the vibrational ground state, 
without the need for a trap operating in the Lamb-Dicke limit. This requires
sideband cooling in a sub-recoil regime. We present a thorough
analysis of sideband cooling on
one or a pair of sidebands simultaneously.
\end{abstract}

\pacs{03.67.Lx, 32.80.Pj, 42.50.Lc}

Laser cooling \cite{Hansch,WD,Ari} and electron shelving detection techniques
\cite{jump,jumprev,Thomp} were developed
in the pursuit of better control of a basic physical system, and
high precision experiments. Up to now such experiments have concentrated
on atomic transitions that offer good prospects as frequency standards,
or that allow sensitive detection of some physical effect. The relatively
new subject of quantum information theory \cite{Ekert,Barenco,Steane} has led 
to interest in a different experimental approach. Here, an important aim is to 
realise the preparation, coherent control, and detailed measurement of some 
sufficiently complex physical system. This allows one to realise simple 
networks of elementary quantum operations (quantum gates), and to test basic 
ideas in quantum information theory, such as data compression, quantum error 
correction, and simple quantum algorithms. The new feature is that such 
experiments do not require high precision as an end in itself
(unlike frequency 
standards), although high precision is of course very desirable, nor do they 
require sensitivity to some physical effect. Instead, one's attention is 
focused purely on the logical properties of the state transformations and 
measurements that can be carried out. 

The significance of experiments on trapped ions to quantum information physics 
was recognised by Cirac and Zoller \cite{CZ}, who proposed the use of a linear 
ion trap to realise the essential elements of a quantum computer. Although a 
real ion trap will not permit large-scale quantum computing 
\cite{Hughes,Stion,WineRev}, such a system should allow experiments on a few 
tens of qubits, and is one of the most promising for such a purpose. Currently 
a few groups worldwide are developing experiments to pursue such
ideas \cite{WineRev}. 

In this paper we consider the question ``what is the simplest method to
build and operate a linear ion trap quantum information processor?''
One of the available options is to use an ion such as calcium or
strontium whose ground state has total spin half. The use of the two
Zeeman sublevels of the ground state of each ion naturally suggests itself
as a means to store each qubit, but currently this option has not
received much attention, partly because the method to measure the
final state, the ``readout'' in computer terminology, is not obvious.
We propose in sections 1 and 2 a simple electron shelving technique that will
permit such measurements. Our method does not realise
an ideal von Neumann measurement (projection onto an orthonormal
basis) but nevertheless permits
efficient quantum computing and quantum state tomography. This is
discussed in section 3.

In section 4 we consider state preparation in this system.
The most difficult part of the state preparation is the cooling of
the motional degrees of freedom. We present the results of a
study of sideband cooling in the regime where the confinement is
of intermediate tightness (Lamb-Dicke parameter of order one), since
this is typically the regime in which one would wish to
operate the trap as a ``processor''. We find that first sideband cooling
operates well for Lamb-Dicke parameters up to $\eta \sim 0.6$, and indeed
sub-recoil cooling is possible. By combining excitation of the first
sideband with a higher one, cooling to the ground state can be achieved for
Lamb-Dicke parameters up to $\eta \sim 3$. We find that the best choice of
sideband is higher than one might expect, and the analysis reveals
the optimum ratio of laser intensities on the two sidebands.

\section{Choice of transition}

Consider a linear ion trap, in which the vibrational frequencies along the 
three axes, for a single confined ion, are $\omega_x \simeq \omega_y \gg 
\omega_z$. Each ion must store one qubit (or possibly
more than one), so we require two long-lived states in the ion,
between which transitions can be driven coherently. The recoil
energy is defined $E_R \equiv (\hbar k)^2 / (2M)$ where $k$ is the
wavevector of the radiation used to drive the transition
and $M$ is the mass of a single ion. The Lamb-Dicke parameter
is
\beq 
  \eta \equiv m \cos(\theta) (E_R / \hbar \omega_z)^{1/2}  \label{eta}
\eeq
where $\theta$ is the angle between the laser beam(s) and the $z$ axis
(${\bf k}\cdot{\bf z} \equiv k \cos(\theta)$) and
$m=1$ for single photon transitions, $m=2$ for Raman transitions,
assuming the geometry ${\bf k}_1 \cdot {\bf z} = - {\bf k}_2 \cdot {\bf z}$.

We require the Lamb Dicke parameter not to be vanishingly small, in order
that the radiation can affect the motional state of the ions, and therefore
optical rather than radio-frequency radiation must be used. 
Since the separation of adjacent trapped ions is in the region
$10$ to $100\;\mu$m, radiation
of sub-micron wavelength is also preferable
in order to address individual ions. These considerations lead to
two approaches to information processing experiments: either
a single-photon transition between metastable states separated
by optical frequencies is adopted, or an rf transition
is driven using optical radiation by means of the stimulated
Raman effect.

We will consider the case of Raman transitions, since the laser
frequency stability requirement is much less restrictive \cite{coolBe}.
The Raman transition is sensitive only to the difference of the
frequency of the two laser beams used to drive it. If both beams
are derived from the same laser, the laser's frequency fluctuations
cancel out, and we only have to concern ourselves (to first order)
with power fluctuations. A single-photon transition, by contrast,
would require a laser system close to the state of the art in
frequency stability.

Atomic energy levels that have long natural lifetimes and
separations in the rf regime almost always owe their energy
separation to the hyperfine interaction, or the Zeeman or
Stark effects. The singly-charged $^9$Be ion is a good candidate for
information processing purposes because it has hyperfine
structure. However, the transition wavelengths in
ions such as $^{40}$Ca and $^{88}$Sr are more easily accessible with
current laser technology. This leads us to consider these ions
as candidates for quantum information experiments. Since they
lack hyperfine structure, and we wish to use long-lived states with
rf separation, we store each qubit in the two-dimensional Hilbert
space spanned by the two Zeeman sublevels $\ket{M=\pm 1/2}_g$
of the ground state. 
The original proposal of Cirac and Zoller
\cite{CZ} made use of a third long-lived state in each ion in order
to bring about quantum gates such as {\sc xor}, which is not
available here. However, we can avoid the
need for such a state either by use of the ``magic Lamb-Dicke parameter''
method of Monroe {\em et al.} \cite{magic}, or by using the first
excited state of the second normal mode of oscillation.

One disadvantage of our choice is that the transition
$\ket{M=-1/2}_g \leftrightarrow \ket{M=+1/2}_g$ is sensitive to the
local magnetic field to first order. Although this would rule out
such a transition as a frequency standard, it does not
for information processing. The reason is that we only
require short term stability, not absolute precision, in the energy
level separations. The actual value of these separations is
totally immaterial, as long as the Raman resonance condition can
be found precisely in any given experiment. An advantage of our choice
is that there are no further sublevels in the ground state into 
which population can ``leak'', which is one source of error in
quantum information processing.

A major apparent problem with our choice is in the final state
detection. In quantum processing, we require a sensitive measurement
of the final state of each individual qubit. 
Sufficient signal to noise ratio to 
make such a measurement is quite rare in physics, but it can be done for 
individual trapped atoms by exciting fluorescence on an optical transition 
that is resonant for one state (say $\ket{M=+1/2}_g$) but
not for the orthogonal 
state. Typically at least a few thousand cycles of excitation followed by 
spontaneous decay must be completed in order to get an unambiguous signal.
This is possible if the optically excited state can only decay to
the $\ket{M=+1/2}_g$ 
ground state, that is not to $\ket{M=-1/2}_g$ and not to some other metastable 
state. However, in ions such as Ca, Sr, Ba and Hg all the short-lived 
excited states (one of which must be used for efficient fluorescence detection)
have a chance of decaying to a metastable state (see fig. 1), 
causing fluorescence to stop after about 3 to 10 cycles. This is the
problem we wish to address. 

\section{Electron shelving Zeeman state measurement}

Our scheme to detect the final Zeeman ground state of a single ion of
Ca, Sr, or a similar candidate, requires three laser wavelengths,
exciting the transitions
$S_{1/2} \leftrightarrow P_{3/2}$,
$S_{1/2} \leftrightarrow P_{1/2}$,
and $D_{3/2} \leftrightarrow P_{1/2}$ (see fig. 1).
Since the latter two are required for laser cooling in any case,
the main cost of the scheme is one additional laser wavelength.
The wavelengths for $^{40}$Ca are 394 nm, 397 nm and 866 nm respectively.
We will use these values as a shorthand to refer to the transitions, although 
obviously the method works essentially unchanged for all similar ions.

A magnetic field is imposed in a direction along the wavevector of the the 
394~nm radiation, which typically is not the $z$ axis of the ion trap.
We adopt 
the magnetic field direction as the quantization axis for the purpose of 
defining spin states and polarizations. The 394~nm radiation is polarized 
$\sigma^+$. The polarization of the other beams is almost unconstrained, linear 
polarization at right angles to the magnetic field direction 
(i.e. equal amounts of $\sigma^+, \sigma^-$) will do.

During the first part of the detection scheme,
the magnetic field must be large enough to enable the Zeeman components of the 
$S_{1/2} \leftrightarrow P_{3/2}$ transition (394 nm) to be resolved. In other 
words, their splitting is of the order of or greater than the inverse of the 
lifetime of the excited state ($P_{3/2}$). The 394~nm radiation is tuned to
resonance with the $S_{1/2}, M=+1/2 \leftrightarrow P_{3/2}, M=3/2$
transition, and therefore excites transitions from $\ket{M=+1/2}_g$
much more strongly than from $\ket{M=-1/2}_g$. 

Let $\Pi_{\pm}$ be the initial populations of the ground state sublevels,
i.e. the diagonal elements of the single-ion density matrix in the 
$\ket{M=\pm 1/2}_g$ basis. 

The measurement proceeds in two steps: shelving followed by fluorescence.
First the 394~nm radiation is pulsed on for a time sufficient to
cause about 10 cycles of excitation and spontaneous decay.
After this time the population that was originally in $\ket{M=+1/2}_g$
has been optically pumped (shelved) into the $D$ states, most of it
(about 90\%) in the $D_{5/2}$ state, while the population in
$\ket{M=-1/2}_g$ is almost unaffected. For a single ion experiment,
this means that the state jumps abruptly to the $D$ manifold or
to $\ket{M=-1/2}_g$ with probabilities proportional to $\Pi_+$,
$\Pi_{-}$ respectively.
Next, the 394~nm radiation is
turned off, the magnetic field is reduced,
and the two other wavelengths are turned on. Fluorescence
is detected during a time equal to a few percent of the lifetime
of the $D_{5/2}$ state. Since this time is more than $10^7$
times longer than that of the $P_{1/2}$ state, of the order of
a million cycles of fluorescence can take place, making a large
accumulated signal. The probability of observing fluorescence
is proportional approximately to
$\Pi_{-} + 0.1 \Pi_{+}$. A further signal for normalisation
or checking purposes can be obtained after the $D_{5/2}$ state
decays, if the ion state jumped there during the shelving pulse.

We have analysed this scheme by numerically solving the complete
set of rate equations, including the 18 Zeeman sublevels
and 40 allowed electric dipole transitions in 
the problem. This enables us to calculate the requirements on magnetic
field strength and laser pulse timing, and to interpret the signal
precisely. A rate equation rather than optical Bloch equation approach
is sufficient since during the shelving pulse, optical coherences play no role 
in determining populations, and during the fluorescence detection we impose a 
magnetic field sufficient to prevent accumulation of population in a
``dark state'', that is, a spin state in the $D_{3/2}$ Zeeman manifold that 
does not couple to the 866 nm radiation. For this, the dark state must Larmor 
precess to a bright state much faster than the optical pumping rate among the 
Zeeman sublevels of the $D_{3/2}$ state. The calculation tells us this optical 
pumping rate.

In order to write the equations, let us number the five energy levels
in order of increasing energy, as in fig. 1. Let $N^{(n)}_M$ be the
population of the $M$ Zeeman sublevel of the $n$th level. 
The total rate of change of population $N^{(n)}_M$ is
$(dN^{(n)}_M/dt)_{\rm sp} + (dN^{(n)}_M/dt)_{\rm dr}$,
a sum of spontaneous and driven terms. The spontaneous feeding
and decay terms are
  \beq
\left( \frac{d N^{(n)}_M}{dt} \right)_{\rm sp}
= \sum_{n' > n} \Gamma_{nn'} \sum_{q=-1}^1
\left| C_{J(n),M}^q \right|^2 N^{(n')}_{M+q}
- \sum_{n' < n} \Gamma_{n'n} N^{(n)}_M,          \label{dNsp}
  \eeq
where $C_{J(n),M}^q $ are the Clebsch-Gordan coefficients.

The driven terms are optical excitations caused by the radiation field,
taking into account the Rabi frequency $\Omega$, detuning $\delta$ and
the polarisation of the radiation. For example, during the
initial ``shelving'' stage when only $\sigma^+$ polarized
394~nm radiation is present, tuned to resonance with the 
Zeeman-shifted $S_{1/2}, M=1/2 \leftrightarrow P_{3/2}, M=3/2$ transition,
the driven terms are
  \begin{eqnarray}
\left( \frac{dN^{(1)}_{1/2}}{dt} \right)_{\rm dr} &=&
- \left( \frac{dN^{(5)}_{3/2}}{dt} \right)_{\rm dr}
=
-\left( N^{(1)}_{1/2}  - N^{(5)}_{3/2} \right)
\frac{\pi}{2} \Omega_{15}^2 \, g_5(0) , \\
\left( \frac{dN^{(1)}_{-1/2}}{dt} \right)_{\rm dr} &=&
- \left( \frac{dN^{(5)}_{1/2}}{dt} \right)_{\rm dr}
- \left( N^{(1)}_{-1/2}  - N^{(5)}_{1/2} \right)
\frac{\pi}{2} \frac{1}{3} \Omega_{15}^2 \, g_5 \!\left(
\mbox{\small $\frac{2}{3}$} \mu_B B / \hbar \right),
  \end{eqnarray}
where the lineshape function is defined by
  \begin{eqnarray}
g_i \!\left( \delta \right) &=&
\frac{ \Gamma_i / 2 \pi }{\delta^2 + \Gamma_i^2/4},   \label{g} \\
\Gamma_i &=& \sum_j \Gamma_{ji}.
  \end{eqnarray}
We obtain the spontaneous decay rates $\Gamma_{ji}$ from \cite{decayrates}.
For Ca, they are
$\Gamma_{14} = 2\pi \times 20.7$ MHz,
$\Gamma_{24} = 2\pi \times 1.69$ MHz,
$\Gamma_{15} = 2\pi \times 21.5$ MHz,
$\Gamma_{25} = 2\pi \times 0.177$ MHz,
$\Gamma_{35} = 2\pi \times 1.58$ MHz. The Clebsch-Gordan coefficients
and Land\'e $g$ factors are obtained by calculation in the $LS$ coupling
approximation. Note that we have neglected the far-off-resonant excitation
rates in the rate equations written above. They were in fact included in the
numerical calculations, though their influence is negligible.

We can quantify the success of the spin measurement by two parameters:
the ability to discriminate the $\ket{M=\pm 1/2}_g$ states, and the
size of the signal. We parametrize the former by calculating a
``shelving efficiency'' $\epsilon$ defined to be
$\epsilon = S_+ - S_-$, where $S_+ = \sum_M N^{(3)}_M$ is
the population in the $D_{5/2}$ levels after the shelving pulse,
when the population is initially all in
$\ket{M=+1/2}_g$ (i.e. $\Pi_{+} = 1$), and $S_-$ is the same quantity
when the population is initially all in $\ket{M=-1/2}_g$ (so $\Pi_-=1$).
In fig. 2 we show $\epsilon$ as a function of the magnetic field $B$,
for two values of the Rabi frequency, $\Omega_{15} = \Gamma_{15}$ (dashed
line)
and $\Omega_{15} = \Gamma_{15}/4$ (full line), with the shelving pulse
duration optimized for maximum $\epsilon$ at each value of $B$.
At large fields, $\epsilon$ tends to $\Gamma_{35} / (\Gamma_{35}
+ \Gamma_{25}) \simeq 0.899$. The field required to obtain
$\epsilon > 0.5$ is of order $B \simeq 3 \times 10^{-3}$ T, which is easily 
produced in the lab. 

It is important to get the shelving pulse duration right, since
if it is too long then all the ground state population will be
pumped to the $D$ states, irrespective of the initial
values of $\Pi_{\pm}$. This duration, and the required magnetic field,
can be estimated by a reduced set of rate equations, in the limit
that all rates are fast compared to two, namely the rate $R_1$ of
transfer from $\ket{M=-1/2}_g$ to $\ket{M=+1/2}_g$, and the
rate $R_2$ of transfer from $\ket{M=+1/2}_g$ to the $D$ states.
These are
  \begin{eqnarray}
R_1 &=& \frac{2}{3}
\frac{\pi}{2} \frac{1}{3} \Omega_{15}^2 \, g_5 \!\left(
\mbox{\small $\frac{2}{3}$} \mu_B B / \hbar \right),\\
R_2 &=& \frac{ \Omega_{15}^2 }{ 2 \Omega_{15}^2 + \Gamma_5^2 }
\Gamma_{35}.
  \end{eqnarray}
The reduced set of equations are the following:
  \begin{eqnarray}
\frac{dN^{(1)}_{-1/2} }{dt} &=& -R_1 N^{(1)}_{-1/2}, \\
\frac{dN^{(1)}_{1/2} }{dt} &=& R_1 N^{(1)}_{-1/2} - R_2 N^{(1)}_{1/2}, \\
\frac{dN^{(D)}}{dt}        &=& R_2 N^{(1)}_{1/2},     \label{dND}
  \end{eqnarray}
where the final equation accounts for all population in the $D$ levels.
Writing $S_{+} = 0.899 N^{(D)}(t)$ when $N^{(1)}_{1/2}(0) = 1$,
$S_- = 0.899 N^{(D)}(t)$ when $N^{(1)}_{-1/2}(0) = 1$, we obtain the
solution
  \beq
\epsilon (t) = 0.899 \frac{R_2}{R_2 - R_1}
\left( e^{-R_1 t} - e^{-R_2 t} \right).  \label{epst}
  \eeq
This solution reproduces the result of the complete calculation quite
well, see fig. 3.
The optimum pulse length, when $d\epsilon/dt = 0$, is therefore
  \beq
t_{\rm max} = \frac{ \ln (R_2 / R_1) }{R_2 - R_1},
  \eeq
and the requirement on the magnetic field is $R_2 \gg R_1$
which is roughly equivalent to the requirement 
that the Zeeman splitting is large compared
to $\Gamma_5$, as one would expect. The typical order of magnitude
for $t_{\rm max}$ is $10/\Gamma_{35} \simeq 1\;\mu$s. 

To complete the analysis,
we also solved the complete set of rate equations during the detection
phase, when laser radiation at 397 nm and 866 nm is present
(polarised perpendicular to the magnetic field direction). 
The observed fluorescence rate will be proportional to the population
of the $P_{1/2}$ level, $N^{(4)}_{-1/2} + N^{(4)}_{+1/2}$.
We show in fig. 4 the steady state value of this
population as a function of magnetic field, with the 
lasers always tuned to the zero-field resonance frequencies, and 
initial condition $\Pi_- = 1$. 
Comparing fig. 4 with fig. 2, the conclusion is that it is not
absolutely necessary to reduce the magnetic field
between the shelving and detection phases, as long as sufficient laser power
is used during the detection phase, but that a reduction to $10^{-3}$ T is
useful to increase the observed signal. Since the
$D_{5/2}$ state lifetime is long, there is plenty of time to reduce the 
magnetic field if so desired. 

We find also that the steady-state $P_{1/2}$ population remains of order $1/4$ 
when the magnetic field is large enough to produce Larmor precession much
faster than the optical pumping rate among the Zeeman sublevels
of $D_{3/2}$. This is 
made possible by the favourable condition $\Gamma_{14} \gg \Gamma_{24}$,
and is necessary in order to avoid population
accumulating in a dark state of $D_{3/2}$. 

\section{Requirements on measurements for quantum computing}

The scheme we have discussed does not achieve an ideal measurement, which
would be represented by the value $\epsilon = 1$. In this section we will
discuss whether this is a problem.

First consider measuring a single ion. Any $\epsilon > 0$ will allow the
populations $\Pi_{\pm}$ to be deduced by repeating the measurement many
times (preparing the same state each time). It is sufficient that the
graph of expected signal (the mean number of times fluorescence is seen
during the detection phase) verses $\Pi_{+}$ is single-valued, since then
$\Pi_+$ (and therefore $\Pi_-$) can be deduced unambiguously from the
mean signal. If $\epsilon$ is small the measurement will need to be repeated
more times in order to get good statistics to estimate $\Pi_+$, but our
value $\epsilon \simeq 0.9$ is not significantly worse than $\epsilon = 1$
in this respect. Methods available for other ions, such as probing a 
cycling transition in Be, have $\epsilon$ closer to 1, but the signal to noise
ratio is worse because imperfection in the laser polarisation limits the
number of fluorescence cycles per measurement to a few thousand, instead of
about a million as in our method.

Now consider a many-ion quantum processor containing $N$ ions. It is clear
that $\epsilon \sim 0.9$ is sufficient for information processors of small
or intermediate size, simply because $\epsilon^N$ remains non-negligible
(it falls to $0.01$ at $N=44$). The question is whether such non-ideal
measurements remain useful for a large quantum computer, containing 
possibly thousands of qubits. Although ion traps may not offer the best
technology to build large quantum computers in the future, the question
of measurement errors will remain significant. We will discuss
this using the language of ion traps; the results can be
translated directly to other systems.

In the context of quantum computation, the final state of the
computer represents the output of some algorithm. First consider the
case that this output (before measurement) is a product
state where each ion is either in the
state $\ket{M=-1/2}_g$ or $\ket{M=+1/2}_g$. The probability
of obtaining a correct readout after a single run of the algorithm
is $\epsilon^N$ where $N$ is the number of ions in $\ket{M=+1/2}_g$.
It might be imagined that this exponential dependence on $N$ renders
the complete algorithm (including final measurement)
computationally inefficient. In fact this is not the case. On repeating
the whole algorithm
and final measurement $r$ times, independent statistics are gathered on the
final state of each ion, with the aim of deciding whether 
the mean number of times fluorescence is obtained is equal to
$r$ or $r (1- \epsilon)$. The probability that this mean result is
interpreted correctly is of order $[1 - (1-\epsilon)^r]^N$. The number
of repetitions required to make the overall success probability close to one
is therefore $r \sim O( \log N / \log (1-\epsilon)^{-1})$. The method
is thus efficient.

In the case that the final state $\ket{\phi_f}$ (before measurement) of
the quantum computer is not a single product state, simply repeating the
whole computation may not succeed since the state $\ket{\phi_f} =
\sum_u a_u \ket{u}$ can consist of a superposition of product states $\ket{u}$
in which the number of terms in the superposition is exponentially large,
as for example in Shor's algorithm. In this case the problem can be avoided
by error-correction coding \cite{StRS}, using a repetition code in
the measurement basis. The coding requires a repetition $r$ of the same order
as that given in the previous paragraph. The measurement of the computer's
final state must be post-processed (by classical computation) to implement
the error correction, after which the value $u$ for one of the product
states $\ket{u}$ in $\ket{\phi_f}$ can be deduced with high probability.
The situation is now equivalent to having ideal measurements.

Large quantum computers will rely heavily on quantum error correction, applied
repeatedly during the computation. Such fault-tolerant methods work best when
the error syndromes can be measured, rather than handled by unitary
quantum networks \cite{Space}, since the non-trivial syndrome interpretation
can then be carried out reliably by a classical computer. Hence it is
important to ask whether $\epsilon \sim 0.9$ renders the syndrome measurement
too unreliable. A preliminary calculation using the methods of fault-tolerant
syndrome extraction suggests that reliable computing can be achieved with
this value of $\epsilon$, however this is a long calculation and will be
presented elsewhere. It is a helpful feature that the zero syndrome,
corresponding to no error, can be measured reliably.

\section{Sideband cooling}

So far we have established that quantum information
processing experiments can be carried out in the ground state of
ions such as $^{40}$Ca$^+$, with good final readout. The choice of
state avoids the need for very highly stabilised lasers, and 
the Zeeman shelving method avoids the need for
very efficient fluorescence detection, making such experiments
significantly simpler than they would be otherwise. In this
section we consider a further simplication, in which the
need for very tight confinement in the ion trap is avoided.

Cooling of the motional degrees of freedom is one of the most
difficult parts of quantum processing experiments. So far
only single ions have been cooled to the quantum ground state of
a trap, first in one dimension only \cite{Died89}, then in three
dimensions \cite{coolBe}. Both experiments used {\em sideband cooling}
\cite{WD,Neu78} in the Lamb-Dicke limit. The Lamb-Dicke limit is the
condition that the ion's motion is confined to a region small
compared to the wavelength of the radiation under consideration.
Since we are concerned with motional states near
the ground state, of dimension $a_0$, 
this condition is $\eta = 2 \pi a_0/\lambda \ll 1$. From equation
(\ref{eta}), this places a constraint on the confinement provided
by the trap, parametrised by the vibrational frequency along
the $z$ axis, $\omega_z \gg E_R / \hbar$. Single ions of the species we are
considering have recoil frequencies $E_R / h$ in the region 6 to
30 kHz. Taking as an example a string of 10 ions we
then require $\omega_z \sim 2 \pi \times 300$ kHz. 
To make the ions lie along a line, we ned also
$\omega_z \ll \omega_{x,y} \sim 2\pi \times 3$ MHz. This degree of confinement 
is experimentally accessible, but requires either
very small closely spaced trap electrodes, or careful high voltage
r.f. design. In the former case the coupling between the
ion motion and noise voltages in the electrodes is enhanced, and in the
latter the r.f. voltages are liable to be more noisy. Therefore there
are advantages in using a less tightly confining trap.

Another reason to avoid the Lamb-Dicke limit is that we need to drive
vibrational-state-changing transitions ($\Delta n = \pm 1$, where
$n$ is the vibrational quantum number of the fundamental vibrational
mode) using the laser radiation, for
information processing in the ions. A small value of $\eta$
is not optimal for this purpose because it increases the
off-resonant excitation of $\Delta n = 0$ transitions when
$\Delta n = \pm 1$ transition are driven. One could
achieve the right conditions by first cooling to the ground state in the 
Lamb-Dicke limit, and then adiabatically opening the trap to $\eta 
\sim 1$ before processing, but it is interesting to know if this adiabatic 
opening can be avoided. Furthermore, for $\eta \simeq 1$ the cooling rate is 
higher than for $\eta \ll 1$, which permits a lower
steady-state temperature to be reached since heating effects (other
than the random recoil from spontaneous emission) typically
scale more slowly with trap tightness than the sideband cooling rate.

Our approach to sideband cooling in the case of intermediate confinement,
i.e. $\eta \sim 1$, is essentially to use both the first and a higher order
sideband. 
Second sideband cooling in a standing wave was considered by
de Matos Filho and Vogel \cite{Vogel}, who showed that nonclassical
motion can be obtained. We consider travelling wave radiation and are
interested merely in obtaining low temperatures. Morigi {\em et al.}
\cite{Morigi} recently discussed similar ideas to ours. They considered
both $\eta \simeq 1$ and the case $\eta > 3$ where new features
become apparent. We provide a more complete analysis of the region
$0 < \eta < 2$. Other theoretical analyses
\cite{Wine79,Lind84,Sten86,Wine87,Jav93} have been restricted to
first-sideband cooling and the Lamb-Dicke limit, though typically with
a more thorough description of the atom-light interaction or the trapping
potential or both.

To describe the atom-laser interaction in the case of Raman transitions we 
adopt the standard theoretical device of considering the ground state manifold 
$\ket{M=\pm 1/2}_g$ as an effective two-level atomic
system \cite{Mol92,Marz94}. This 
approximation is valid when the excited state populations are negligible. The 
effective Rabi frequency describing the atom--laser coupling is $\Omega = 
\Omega_- \Omega_+ / 2 \Delta$ where $\Omega_{\pm}$ are the single-photon Rabi 
frequencies for excitation from $\ket{M=\pm 1/2}_g$ to an excited
state such as $P_{1/2}$ or $P_{3/2}$ (including Clebsch-Gordan
coefficients), and $\Delta \gg 
\Omega_{\pm}$ is the detuning from resonance of these transitions. The 
effective detuning is 
  \beq
\delta = \omega_{L-} - \omega_{L+} - \omega_{+-} - 
\frac{ \left| \Omega_-\right| ^2 - \left| \Omega_+\right| ^2}
{4 \Delta},
  \eeq
where $\omega_{L\pm}$ are the laser 
frequencies and $\omega_{+-} = \mu_B B/\hbar$ is the Zeeman splitting of the 
$\ket{M=\pm 1/2}_g$ levels. The effective linewidth $\Gamma$ is equal to the 
optical pumping rate from $\ket{M=+1/2}_g$ to $\ket{M=-1/2}_g$ caused by a 
$\sigma^-$-polarized laser resonant with a transition such as $\ket{M=+1/2}_g 
\leftrightarrow \ket{P_{1/2}, M=-1/2}$. This laser radiation is either very 
weak and continuously present, or moderately strong and pulsed on when a 
spontaneous transition is desired. 

\subsection{First sideband cooling}

We will discuss cooling of a single ion along one dimension.
The generalisation to many ions and three dimensions is
straightforward. Following Wineland and Itano \cite{Wine79}, we
will first account for the average change in total energy of 
the ion during one complete cycle of `excitation' ($\ket{M=-1/2}_g \rightarrow 
\ket{M=+1/2}_g$) followed by `spontaneous emission' ($\ket{M=+1/2}_g 
\rightarrow \ket{M=-1/2}_g$). The average is taken over many such cycles, in 
which the direction of spontaneously emitted photons varies randomly.
The total state of the ion will be written
$\ket{M=\pm 1/2}_g \ket{n}$
where the $\pm$ indicates the internal state, and the integer $n \ge 0$
indicates the vibrational state (external energy eigenstate) of the
ion in the trap. 

Let
  \beq
I_{fn} \equiv \left| \bra{f} e^{i \bf k \cdot z} \ket{n} \right|^2.
\label{Ifn}
  \eeq
where $z$ is the position of the centre of mass of the ion. 
This is the relative strength of the different sideband
components in the ion-radiation field interaction \cite{Wine79}. Using
${\bf k \cdot z} = \eta(\hat{a}^{\dagger} + \hat{a})$ where
$\hat{a} \ket{n} = \sqrt{n} \ket{n-1}$ one obtains
  \beq
\bra{f} e^{i \bf k \cdot z} \ket{n} =
e^{-\eta^2/2} \sqrt{n! f!} (i\eta)^{|f-n|} \sum_{m=0}^{{\rm min}(n,f)}
\frac{ (-1)^m \eta^{2 m} }
{m! (m+|f-n|)! ( {\rm min}(n,f) - m )! } .  \label{kz}
  \eeq

The optical cross section for the `absorption'
transition $\ket{M=-1/2}_g\ket{n} \rightarrow \ket{M=+1/2}_g \ket{f}$ is
  \beq
\sigma_{fn} (\delta) = \sigma_0 I_{fn}
\tilde{g} \left( \delta - (E_f - E_n)/\hbar \right)
  \eeq
where $\tilde{g} = g \Gamma \pi/2$ is the lineshape function (\ref{g})
normalised such that $\tilde{g}(0) = 1$. 
If the occupation probability of vibrational state $n$ is $P_n$,
then the rate of change in total energy of the ion during absorption,
averaged over many cycles, is 
  \beq
\frac{d E_{\rm abs}}{dt} = \sum_n P_n \sum_f 
\left( E_f - E_n \right) (I/\hbar \omega_{+-}) \sigma_{fn}, \label{Eabs}
  \eeq
where $I$ is the effective laser intensity.

The average change in ion energy during spontaneous emission is
  \begin{eqnarray*}
E_{\rm sp} &=& \int d \Omega_{\bf k} P({\bf k}) \sum_n
\left( E_n - E_f \right) I_{nf}  \\
&=& \int d \Omega_{\bf k} P({\bf k}) \sum_n
\bra{f} e^{i \bf k \cdot z} \ket{n} \bra{n}
[ H^{\rm ext}, e^{-i \bf k \cdot z} ] \ket{f}  \\
&=& \int d \Omega_{\bf k} P({\bf k}) 
\bra{f} \frac{\hbar^2 k^2}{2M} - \frac{\hbar {\bf k}}{M} 
\cdot {\bf p} \ket{f} \\
&=& E_R.
  \end{eqnarray*}
Note that since $\bra{f} {\bf p} \ket{f} = 0$ 
this answer is independent of the distribution $P({\bf k})$
of directions of spontaneous emission, which is counterintuitive.
If, instead, the absorption and spontaneous
emission are treated as a single scattering process 
(see eq. (\ref{Gfn})), then the same result for the mean energy change
is obtained as long as the distribution $P({\bf k})$ is symmetric about
the origin, as it is in practice.

Combining the contributions of absorption and spontaneous emission, the
mean rate of change of total energy, per cycle of absorption followed
by spontaneous emission, is
  \beq
\frac{d \left< E \right>}{dt} = \frac{I \sigma_0}{\hbar \omega_{+-}}
\sum_n P_n \sum_f \left( E_f - E_n + E_R \right) I_{fn}
\tilde{g} \left( \delta - (E_f - E_n)/\hbar \right)   \label{dEdt}
  \eeq
This equation was derived in \cite{Wine79} as part of a more general
discussion. The above shows that its derivation can be simple and
physically intuitive. It is valid to separate
absorption and spontaneous emission because typically the experimental
procedure is to switch on the Raman lasers and optical pumping laser
at separate times \cite{Died89,coolBe}. However, later we will not
separate them but treat the emission and absorption as a
photon scattering process.

Although we are interested in the case $\eta \sim 1$, it is instructive
first to examine the Lamb-Dicke limit $\eta \ll 1$. This is done
by a series expansion of eq. (\ref{dEdt}) in powers of $\eta^2$. The
expansion of $I_{fn}$ (eq. (\ref{Ifn})) to order $\eta^4$ is
  \begin{eqnarray*}
I_{fn} &=& \delta_{f,n} \left(1 - \eta^2 (2n+1) + \frac{\eta^4}{2}
(3n^2 + 3n + 1) \right) \\
&& + \delta_{f,n-1} \eta^2 n(1- \eta^2 n) + \delta_{f,n+1} \eta^2 (n+1)
(1 - \eta^2 (n+1))\\
&& + \delta_{f,n-2} \frac{\eta^4}{4} n(n-1) + \delta_{f,n+2} \frac{\eta^4}{4}
(n+1)(n+2).
  \end{eqnarray*}
This can be obtained either from eq. (\ref{kz}) or by expanding the
operator $e^{i \bf k \cdot z}$ in eq. (\ref{Ifn}). The terms in eq.
(\ref{dEdt}) up to order $\eta^2$ are
  \begin{eqnarray}
\frac{d \left< E \right>}{dt} &=&
\frac{I \sigma_0 E_R}{\hbar \omega_{+-}} \sum_n P_n \left\{
( \tg_m + \tg_{m+1}) + \eta^2( \tg_{m+2} - \tg_m ) \right. \nonumber \\
&& + n \left(
(\tg_{m+1} - \tg_{m-1}) + \eta^2(\tg_{m-1} - \tg_{m+1} + \tg_{m-2}/2
- 2 \tg_m + 3\tg_{m+2}/2) \right) \nonumber \\
&& \left. + n^2 \eta^2(-\tg_{m-2}/2 - \tg_{m+1} + \tg_{m-1} + \tg_{m+2}/2)
\right\},    \label{dE2}
  \end{eqnarray}
where $m=1,2,\ldots$ indicates the sideband chosen for cooling, and
$\tg_m \equiv \tg(m\omega_z) \equiv \Gamma^2/(4 m^2 \omega_z^2 + \Gamma^2)$.

For first sideband cooling, $m=1$, in the limit of well-resolved sidebands
we ignore terms of order $\Gamma^2/\omega_z^2$ compared to 1 to obtain,
for the steady state distribution
$(d\langle E\rangle /dt=0)$,
  \beq
\tg_1 + \tg_2 + \eta^2 (\tg_3 - \tg_1) + \left< n \right>
(-1 + \eta^2 ) + \left< n^2 \right> \eta^2 = 0  \label{n1}
  \eeq
In the Lamb-Dicke limit one finds that the rate of driving of population
from one level to another is independent of the level number $n$, so the
principle of detailed balance leads to the thermal distribution
defined by $P_{n+1} / P_n = e^{-\hbar \omega_z/k_B T} \equiv s$.
The populations are then
\beq
P_n = (1-s) s^n
\eeq
and the mean and mean square quantum number are
  \beq
\left< n \right> = \frac{s}{1-s}, \;\;\;\;
\left< n^2 \right> = \frac{s(1+s)}{(1-s)^2}.
  \eeq
from which $\left< n^2 \right> = 2 \left< n \right>^2 +
\left< n \right>.$ Substituting this in (\ref{n1}) and solving for $n$,
ignoring terms of order $\eta^4$ or $\Gamma^4/\omega_z^4$, we obtain
  \beq
\left< n \right> \simeq \frac{\tg_1 + \tg_2 - \eta^2(\tg_1 - \tg_3)}
{1 - 2 \eta^2} \simeq 
\frac{ \left( 5 - 32 \eta^2 / 9 \right) }{16 \left(1-2\eta^2 \right)}
\frac{\Gamma^2}{\omega_z^2}  \label{n1res}
  \eeq
This result is precise in the Lamb-Dicke limit with well-resolved
sidebands. Figure 5 shows $1-P_0$ as a function of
$\eta^2$ as given by (\ref{n1res}) (dashed line) and by a more exact
treatment (full line) which will be described shortly. It is seen that
(\ref{n1res}) indicates the right qualitative behaviour for moderate
values of $\eta$, though it is not precise, owing to a slight departure
from a thermal population distribution in the more exact treatment.
The main conclusion is that first sideband cooling works very well
for values of the Lamb-Dicke parameter up to $\eta \sim 0.6$. It is
noteworthy that the effective temperature given by
$k_B T = \hbar \omega_z / \ln (P_0/P_1)$ is approximately equal to the
recoil limit $k_B T = E_R$ when $\eta = 0.5$, $\Gamma = 0.2 \omega_z$,
and is sub-recoil for smaller linewidths.

\subsection{Second sideband cooling}

For second sideband cooling, $m=2$, eq. (\ref{dE2}) can be usefully
simplified by neglecting terms of order $O(\eta^2 \Gamma^2/\omega_z^2)$. In
other words, we make immediately the assumption of well-resolved
sidebands. The steady state of eq. (\ref{dE2}) is then given by
  \beq
\left< n^2 \right> \frac{\eta^2}{2}
 + \left< n \right> (\tg_1 - \tg_3 - \frac{\eta^2}{2})
- (\tg_2 + \tg_3) = 0.    \label{n2}
  \eeq
There are two interesting regimes in which this result can be examined.
First, for a very tight trap with $\eta^2 \ll \Gamma^2/\omega_z^2$,
the $\eta^2$ terms are neglible and we have
  \beq
\left< n \right> = \frac{\tg_2 + \tg_3}{\tg_1 - \tg_3} \simeq \frac{13}{32}.
\label{n3}
  \eeq
This can be understood as the condition that all the population is driven to
the ground and first excited vibrational levels, but that these are
almost equally populated since the laser drives off-resonant transitions
from each to the other at almost equal rates. 

Secondly, let us consider
a trap that is less tight though the sidebands are still well resolved:
$\Gamma^2/\omega^2 \ll \eta^2 \ll 1$. Since the population in levels
$n>1$ is optically pumped efficiently towards $n=0,1$,
most of the population is in $P_0$ and $P_1$, and therefore $\left< n^2
\right> \simeq P_1 \simeq \left< n \right>$. Using this to express
$\left<n^2\right>$ in terms of $\left< n \right>$ in eq. (\ref{n2}),
the $\eta^2$ terms cancel, so once again the mean quantum number is given
by (\ref{n3}). This reasoning is valid as long as
$\left<n^2 \right> - \left< n \right> \ll \tg_2 + \tg_3$.

In summary, cooling on the second sideband alone does not allow the ground
state population to reach a value close to 1. Equation (\ref{n3})
is nevertheless useful as a further test (in addition to (\ref{n1res})) on
the numerical treatment to be described in the next section.

\subsection{Cooling on two sidebands}

We now turn our attention to $\eta > 0.6$. For single-frequency
laser radiation on the $m$th sideband, the populations $P_n$
can be deduced from a set of rate equations, in which the rate of transfer
of population from level $n$ to level $f$ is given by
  \beq
\Gamma_{fn}(m) = \frac{\Omega^2}{\Gamma} \int_{-1}^1 du N(u)
\left| \sum_{j=0}^{\infty} \frac{ \bra{f} e^{-ik z u} \ket{j}
\bra{j} e^{i \bf k \cdot z} \ket{n} \Gamma/2 }
{ \omega_z(-m-j+n) + i \Gamma / 2 }  \right|^2,  \label{Gfn}
  \eeq
where $N(u)$ is the angular distribution of spontaneous emission. Note that
we are now treating the atom-laser interaction as a scattering process
(still in the low intensity limit $\Omega \ll \Gamma, \omega_z$).
If one wished to calculate the case that absorption and emission are
separate processes (i.e. a driven transition followed by optical
pumping), one would take the modulus square of each term in
the sum over $j$, instead of the modulus square of the whole sum. We find
however that the two cases give almost identical results, so we will
restrict our discussion to the form (\ref{Gfn}).

For dipole radiation in three dimensions one typically uses
$N(u) = \frac{3}{8}(1+u^2)$. To remain consistent with the one-dimensional
approach we have adopted, we take
instead $N(u) = (\delta(u+1) + \delta(u-1))/2$, which corresponds physically
to spontaneously emitted photons propagating along the $z$ axis. We find
that this correctly reproduces the Lamb-Dicke limit behaviour described
in the previous sections.

When $\eta > 0.6$, single frequency radiation will not cool the ion to
the motional ground state. We therefore consider the next simplest
case, namely two laser frequencies,
one tuned to the first sideband, the other to the sideband $m > 1$ (both
are red sidebands, i.e. below the carrier frequency). To allow two laser
frequencies in the population rate equations (\ref{Gfn}), we perform
an incoherent sum, giving for the rate of transfer from $n$ to $f$
  \beq
\Gamma_{fn} = \Gamma_{fn}(m) + \alpha \Gamma_{fn}(1). \label{Gfntot}
  \eeq
This corresponds physically to the case that the radiation fields at
the two frequencies are not simultaneously present, but pulsed on one
after the other, and we consider the behaviour time-averaged over
such switching. The parameter $\alpha$ is the ratio of the two laser
intensities.

The general physical picture of the behaviour is that the
laser on the $m$th sideband pumps the population towards the lowest
$m$ vibrational levels, and cools overall, while the relatively
low intensity first sideband laser has a small heating effect 
overall, but efficiently pumps the population from the lowest levels
into the ground state. The rate of this pumping into the ground state
is approximately $\alpha (\Omega^2/\Gamma) I_{00} I_{01}
= \alpha (\Omega^2/\Gamma) \eta^2 e^{-2\eta^2}$.

In figures 6a and 6b we present the result of numerically extracting the
steady-state populations, for a set of population rate equations using
the rates (\ref{Gfntot}). The figures show the value of $1-P_0$
for $\eta^2$ in the range $0$ to $3$, for two values of the linewidth,
and for $m=1$ to 4. The numerical calculations used a truncated set
of levels, from the ground state $n=0$ to $n=n_{\rm max}$. Even though
the population is almost all in the low levels, it was found that a large
number of levels ($n_{\rm max} = 100$) had to be included in order to
calculate $1-P_0$. This is because for $\eta > 0.6$ the population
distribution has a long oscillating `tail' which falls off as a low
power of $n$. As a result,
the results in figs 6a and 6b have only approximately $5\%$ accuracy
on the parts of the curves which rise steeply as a function of $\eta^2$.

We found by numerical experiment that the ground state population does
not depend strongly on the laser intensity ratio $\alpha$, and that
the optimal value is approximately given by $\alpha = 1/(3\eta^2)$. This
choice was adopted for all the results presented here. 

The main conclusions from figure 6 are as follows. We find that 
in the limit of well-resolved sidebands, $1-P_0$
is roughly proportional to $\Gamma^2 / \omega_z^2$ for the
whole range of $\eta$ values shown. The best sideband to choose, in order
to maximise $P_0$ at given $\eta$, is higher than one might expect.
A naive argument would suggest that since resonant absorption results in a
kinetic energy decrease by $m \omega_z$, and
spontaneous emission results on average
in a kinetic energy increase of $E_R = \eta^2 \omega_z$, the best choice
is $m$ equal to the smallest integer larger than $\eta^2$, ignoring
the small heating effect of the first sideband laser. In fact we find
the optimum $m$ is the smallest integer larger than approximately
$2\eta^2 + 0.5$. 

The poor performance shown in fig. 6
around $\eta^2 = 1.27$ and $\eta^2 = 2$ is due
to zeroes in $I_{fn}$, namely $I_{23}(\eta^2 = 3 - \sqrt{3}) = 0$
and $I_{12}(\eta^2 = 2) = 0$. The first makes the depopulation of
level 3 inefficient when $m>3$, the second makes the
depopulation of level 2 inefficient when $m>2$. The structure in the
$m=3$ curve at $\eta^2 \simeq 1.5,\; 1.3,\; 1.2, \ldots$
is due to nearly coincident zeroes of $I_{n-3,n}$ and
$I_{n-1,n}$, causing population to accumulate at $n=8,\; 9,\; 10, \ldots$
respectively.
Such coincidences only happen when both sidebands are odd, or both even,
which is why similar structure does not appear in fig. 6 for $m=2,4$.

The method of cooling on a pair of sidebands will not work for
large $\eta$, because the first sideband laser cannot efficiently
drive transitions
from level 1 to 0. This was noted by Morigi {\em et al.}
\cite{Morigi}. We can roughly estimate the maximum $\eta$ for which
cooling to the ground state is possible as follows.

The equations (\ref{Gfn}) are too complicated to solve analytically
when $\eta$ is not small. In order to make rough estimates, we make
several approximations.
First approximate $\Gamma_{fn}$ by ignoring the distribution $N(u)$ and
assuming the interference terms in the sum roughly cancel, leaving
  \beq
\Gamma_{fn} \simeq \frac{\Omega^2}{\Gamma} \sum_{j=0}^{\infty}
I_{fj} I_{jn} (\tg_{n-j-m} + \alpha \tg_{n-j-1}).   \label{fn}
  \eeq
We are interested in the case of low temperature, where most of the population
is in the low-lying levels. In the limit that most of the population is
contained in $P_0$ and $P_1$, the ratio $r \equiv P_0 / P_1$ can be
estimated to be $r \simeq \Gamma_{01} / \Gamma_{10}$. If $r$ is large then
$P_0 \simeq 1$, so cooling to the ground state has been achieved, and
  \beq
P_0 \simeq \frac{\Gamma_{01}}{\Gamma_{01} + \Gamma_{10}}.  \label{P0}
  \eeq
We now examine the ratio
  \beq
\frac{\Gamma_{01}}{\Gamma_{10}} \simeq
\frac{ \sum_{j=0}^{\infty} I_{0j} I_{j1}( \tg_{m+j-1} + \alpha \tg_j) }
{ \sum_{j=0}^{\infty} I_{1}I_{j0}( \tg_{m+j} + \alpha \tg_{j+1} ) }.
\label{ratio}
  \eeq
Since $m > 1$, the only resonant term in the two sums is the $\tg_j$ term
in the numerator for $j=0$. Therefore the ratio may be written, in the
limit $\Gamma^2 \ll \omega_z^2$,
  \beq
\frac{ \Gamma_{01}}{\Gamma_{10}} \simeq
\frac{ I_{00} I_{01} + \beta}{\beta} =
\frac{e^{-2\eta^2} \eta^2 + \beta}{\beta}
  \eeq
where $\beta$ is $O(\Gamma^2/\omega_z^2)$ and we used the expression for
$I_{00}$ and $I_{01}$ given by (\ref{kz}). The factor $\alpha$ cancels
provided it is larger than $\sim \Gamma^2 / \omega_z^2$. We estimate
$\beta$ as follows. By definition,
  \begin{eqnarray}
\beta &\simeq& \sum_{j=0}^{\infty} I_{1j} I_{j0} \tg_{j+1} \nonumber \\
&\simeq& \frac{\Gamma^2}{4\omega_z^2} \sum_{j=0}^{\infty}
e^{-\eta^2} \frac{\eta^{2j}}{j!} e^{-\eta^2} \frac{\eta^{2(j-1)}}{j!}
(j-\eta^2)^2 \frac{1}{(j+1)^2} \nonumber \\
&\simeq&  \frac{\Gamma^2}{4\omega_z^2} e^{-2\eta^2} \eta^{-2}
\sum_{j=0}^{\infty} \frac{(\eta^2)^{2j}}{(j!)^2}  \label{beta}
  \end{eqnarray}
where we used (\ref{kz}) for $I_{1j}$ and $I_{j0}$. The last step used
$(j-\eta^2)^2 \simeq (j+1)^2$ which is true for large $j$ and enables
the sum to be performed. We find by numerical calculation that this
overestimates $\beta$ by a factor $\sim \eta^2$. The sum in (\ref{beta})
is equal to the Bessel function $I_0(2\eta^2)$ whose value for
$2\eta^2 > 2$ is
  \beq
I_0(2 \eta^2) = \frac{e^{2\eta^2}}{\eta \sqrt{4\pi}}
\left( 1 + O((4\eta)^{-2}) \right).
  \eeq
Substituting this in (\ref{beta}) we obtain
  \beq
\beta \sim \frac{\Gamma^2}{4 \omega_z^2} \frac{1}{2 \sqrt{\pi}} \eta^{-3}.
\label{betares}  \eeq
The most important feature of this calculation is that whereas $I_{00} I_{01}$
falls exponentially with $\eta^2$, we find that $\beta$ does not, therefore
it is impossible to achieve $I_{00} I_{01} > \beta$ (and hence a
large ground state population) once $\eta$ exceeds a value that
depends insensitively on $\Gamma/\omega_z$. 
In other words the optical pumping from $n=1$ to $n=0$ becomes
exponentially inefficient. We can estimate the limiting value of $\eta$
by setting $I_{00} I_{01} = \beta$, giving
  \beq
\eta^2 \sim \ln \left(2 \omega_z/\Gamma \right) + \frac{7}{4}
\ln \left( \eta^2 \right) + 2. \label{log}
  \eeq
where the power $\eta^{-3}$ in equation (\ref{betares}) has been replaced
by $\eta^{-5}$ in agreement with our numerical calculation. Equation
(\ref{log}) gives a reasonable estimate of the maximum value of $\eta$
that will permit accumulation of population in the ground state
by cooling on the first and a higher sideband. For $\Gamma = 0.1
\omega_z$, the limiting value is $\eta \simeq 3$, in agreement with
the results of \cite{Morigi}.

In conclusion, let us comment on the best choice of experimental
parameters implied by this study. Our electron-shelving measurement
measurement technique has as its major advantage the possibility of a
high signal-to-noise ratio, since fluoresence can be driven for a long
time before an unwanted transition makes the signal ambiguous. This renders
information processing experiments in the ground state of $^{40}$Ca$^{+}$
and similar ions feasible.
Our study of sideband cooling has indicated the best choice of laser
parameters in order to achieve a large population of the ground state of
motion. In order to retain the simplicity of only cooling on a single
sideband, combined with a not too great suppression of vibrational
state-changing transitions, the choice $\eta = 0.5$ is suitable.

\section*{Acknowledgements}

DS acknowledges the support of the EPSRC, JB of the Ecole Polytechnique, and
AMS of The Royal Society and of St Edmund Hall, Oxford.

\newpage

\section*{Figure captions}

Fig. 1. Low lying levels in $^{40}$Ca$^{+}$. The Zeeman structure of
the levels is shown, though the level spacings are not to scale. The
wavelengths and inverse spontaneous rates of the electric-dipole allowed
transitions are given. The circled numerals indicate the numbering of
the levels adopted for equations (\ref{dNsp}) to (\ref{dND}).

Fig. 2. Shelving efficiency $\epsilon$ as a function of magnetic field
in Tesla, for two values of the Rabi frequency of the shelving pulse:
dashed line $\Omega_{15} = \Gamma_{15}$;
full line $\Omega_{15} = \Gamma_{15} / 4$.
The shelving pulse duration
was chosen to maximise $\epsilon$ for each value of magnetic field
and Rabi frequency.

Fig. 3. Shelving efficiency as a function of shelving pulse duration, for
$\Omega_{15} = \Gamma_{15}$, $B = 0.01$ Tesla. Full line: solution
of full rate equations; dashed line: equation (\protect\ref{epst}).

Fig. 4. Steady state population of $P_{1/2}$
level, $N^{(4)}_{-1/2} + N^{(4)}_{+1/2}$, during the detection phase,
as a function of magnetic field. The observed fluorescence signal
is proportional to this population. The two laser beam intensities are
set to give equal Rabi frequencies,
$\Omega \equiv \Omega_{14} = \Omega_{24}$.
Reading from the upper (full) curve to
the lower (dash-dotted) curve, this Rabi frequency
is $\Omega = 10\Gamma_4$,
$5\Gamma_4,\; 2\Gamma_4, \; \Gamma_4$ respectively.

Fig. 5. First sideband cooling.
The figure shows the difference between the ground state
population $P_{0}$ and 1,
as a function of $\eta^2$ for first sideband cooling with
$\Gamma = 0.1 \omega_z$. Full line: numerical solution of population
rate equations using rates $\Gamma_{fn}(1)$ (eq. (\ref{Gfn}));
dashed line: analytic solution (\ref{n1res}) 
in Lamb-Dicke limit assuming a thermal
population distribution, giving $1-P_0 = s \simeq \left< n \right>$. 

Fig. 6. Cooling on two sidebands. The curves show the
difference between $P_0$ and 1,
as a function of $\eta^2$, for cooling on the first sideband and the
$m$th, where $m=1$ (full line), 2 (dashes), 3 (dots), and 4 (dash-dot),
with $\alpha = 1/(3 \eta^2)$ which roughly maximises $P_0$.
(a): $\Gamma = 0.1 \omega_z$; (b): $\Gamma = 0.2 \omega_z$. The
curves are obtained by numerical solution of the steady-state population
rate equations with 100 vibrational levels included in the calculation.

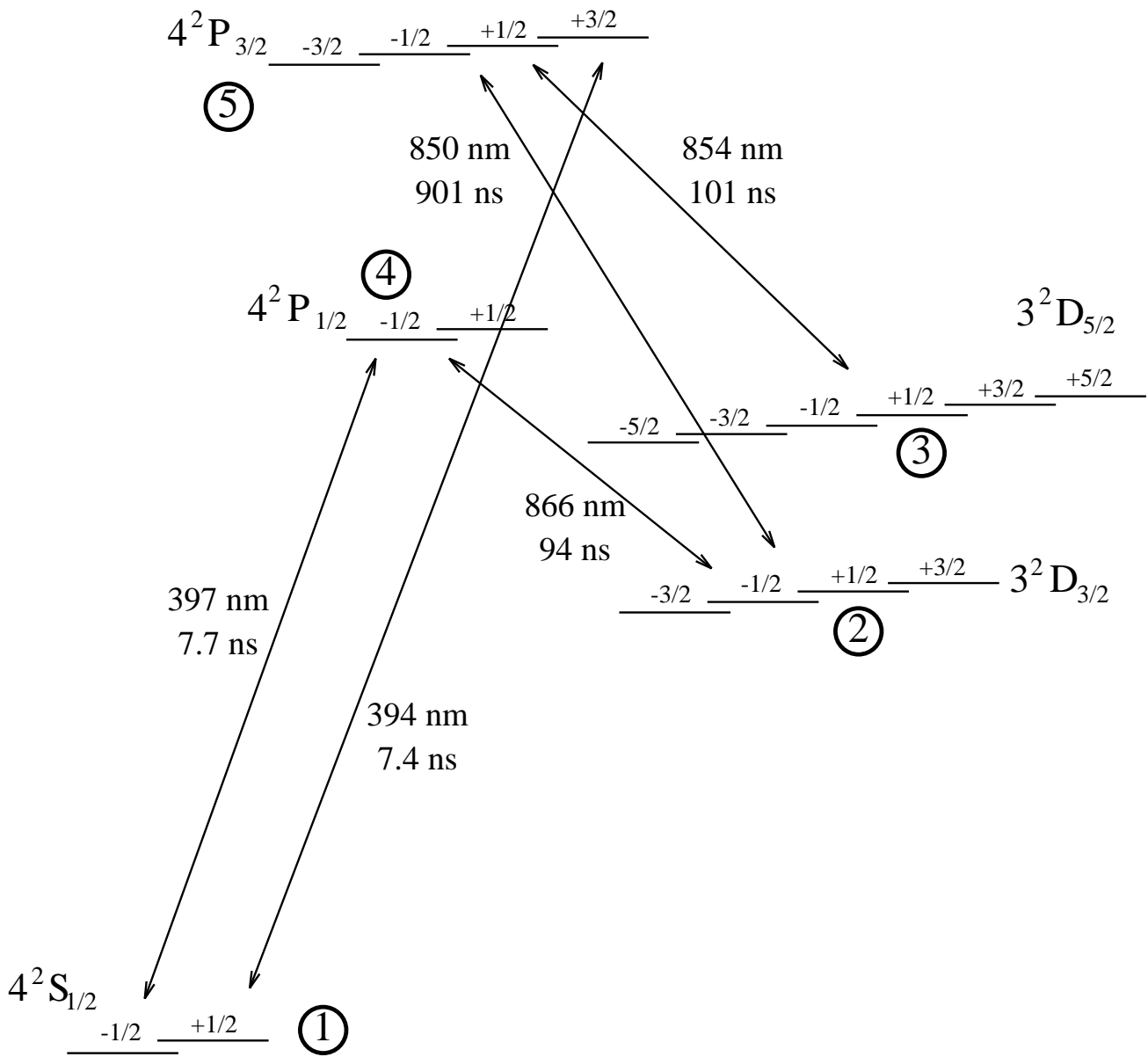
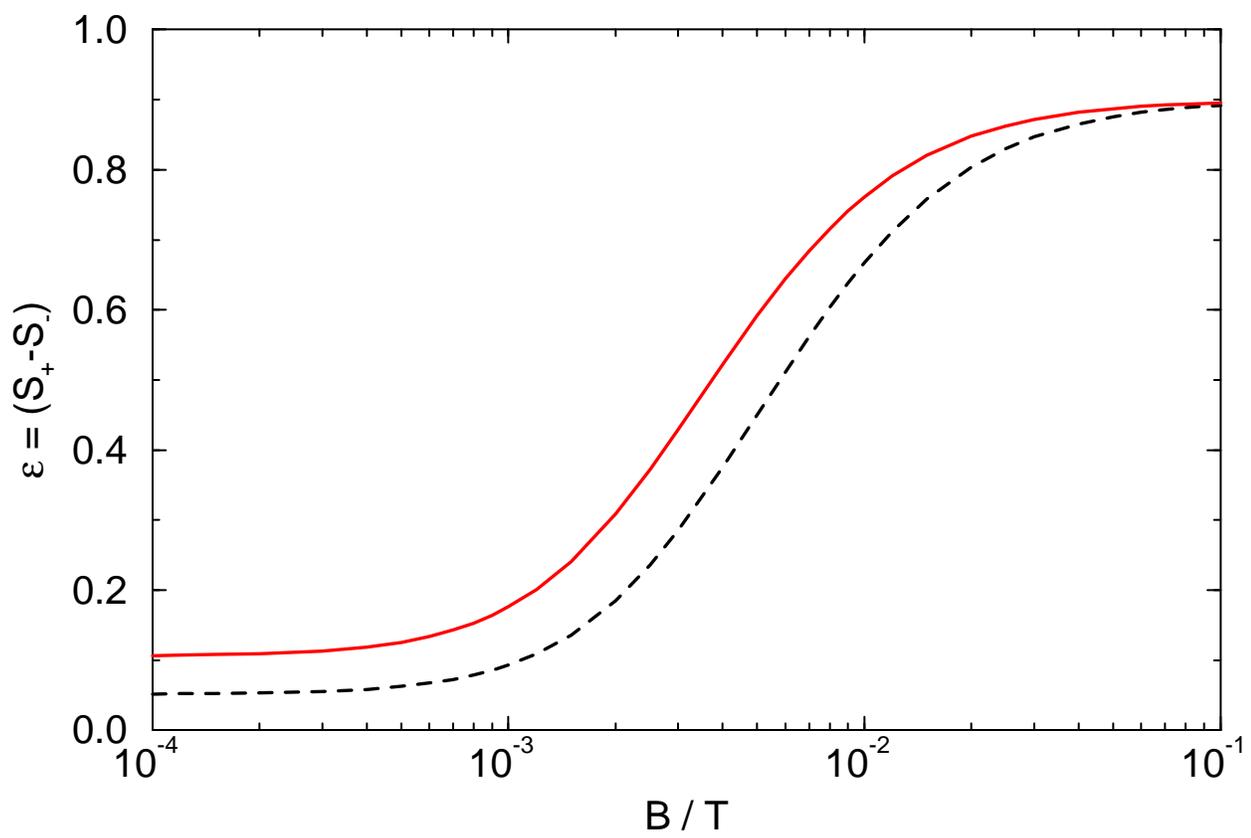
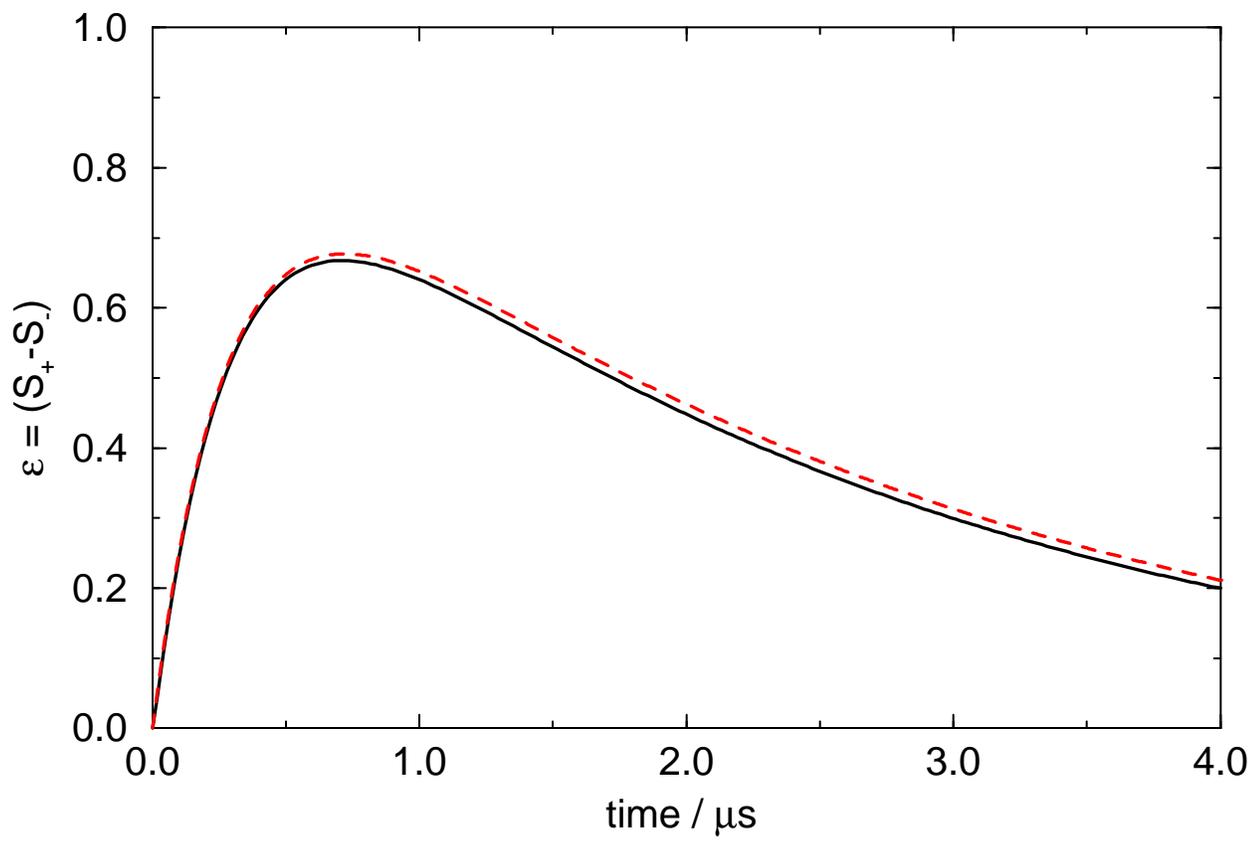
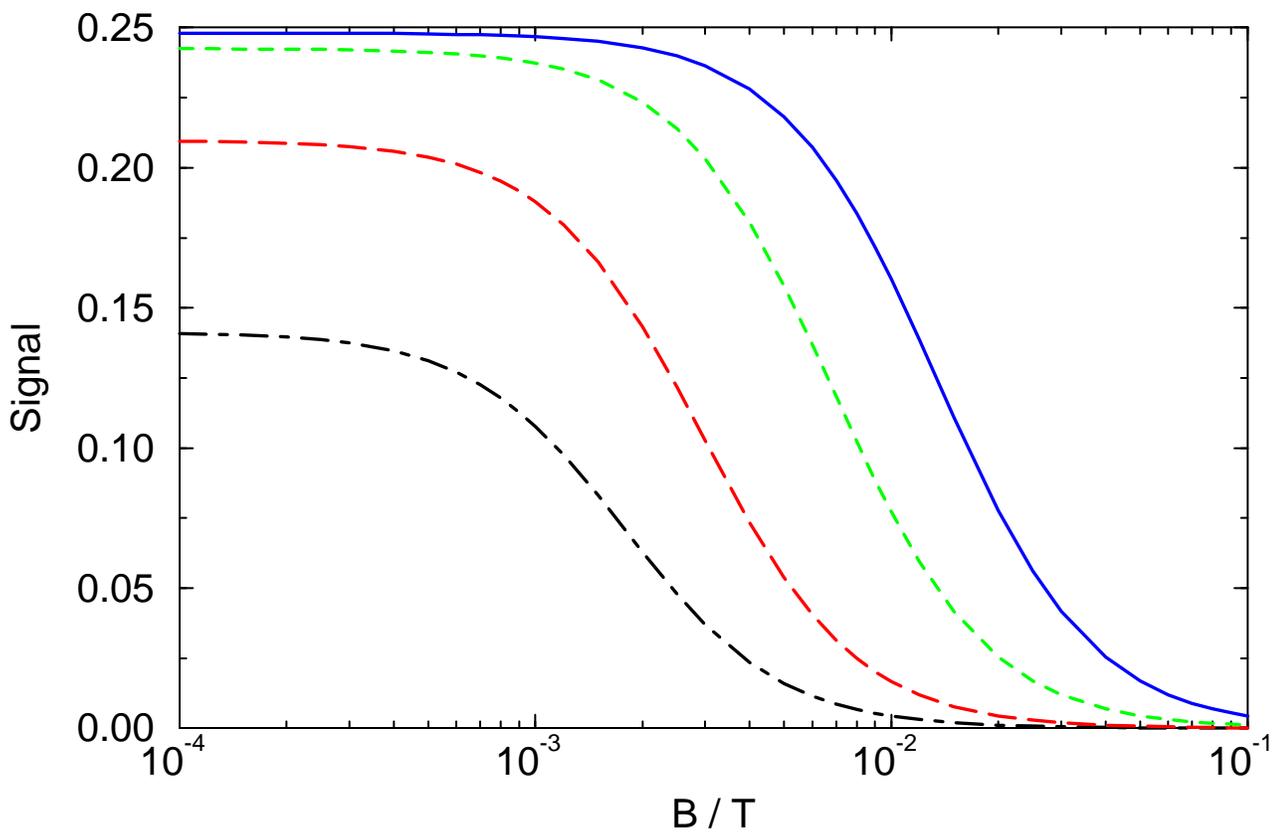
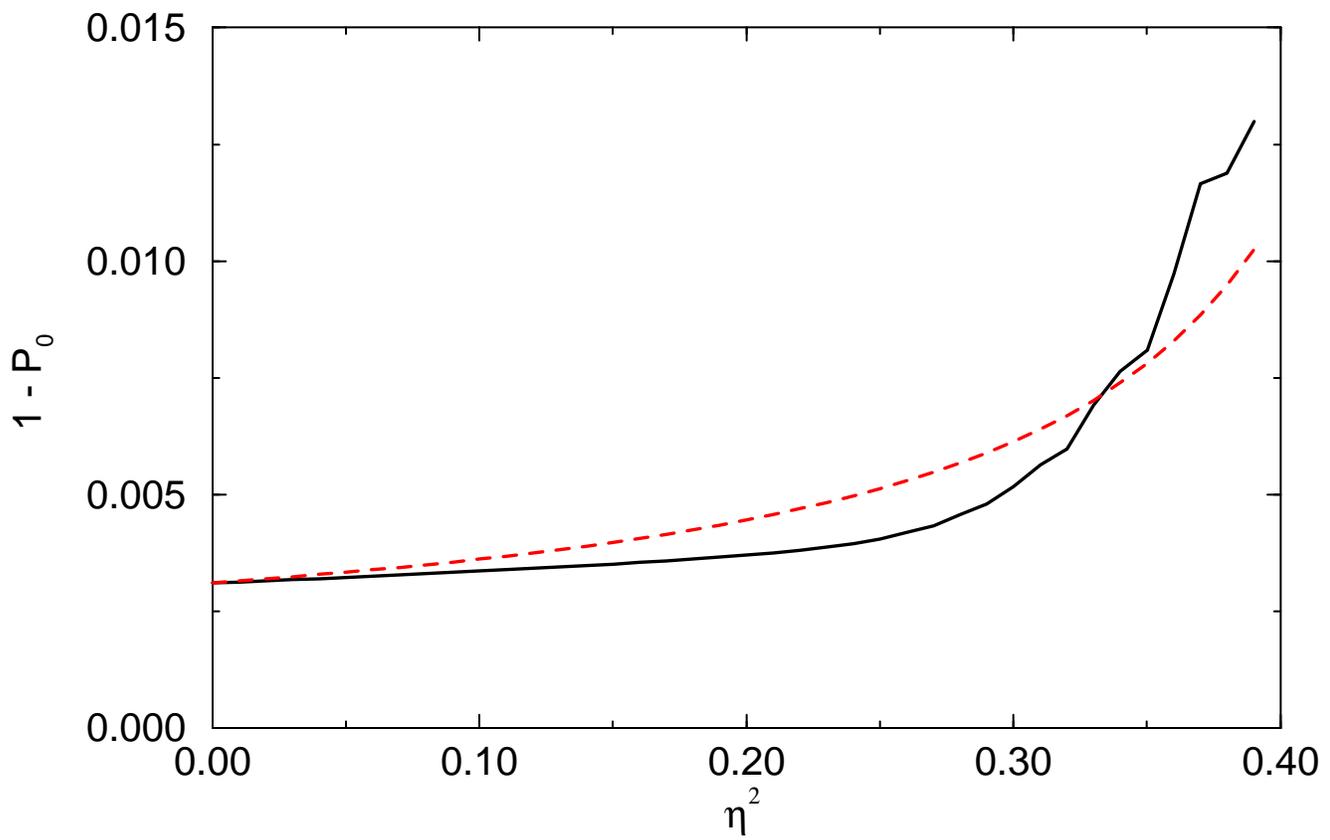
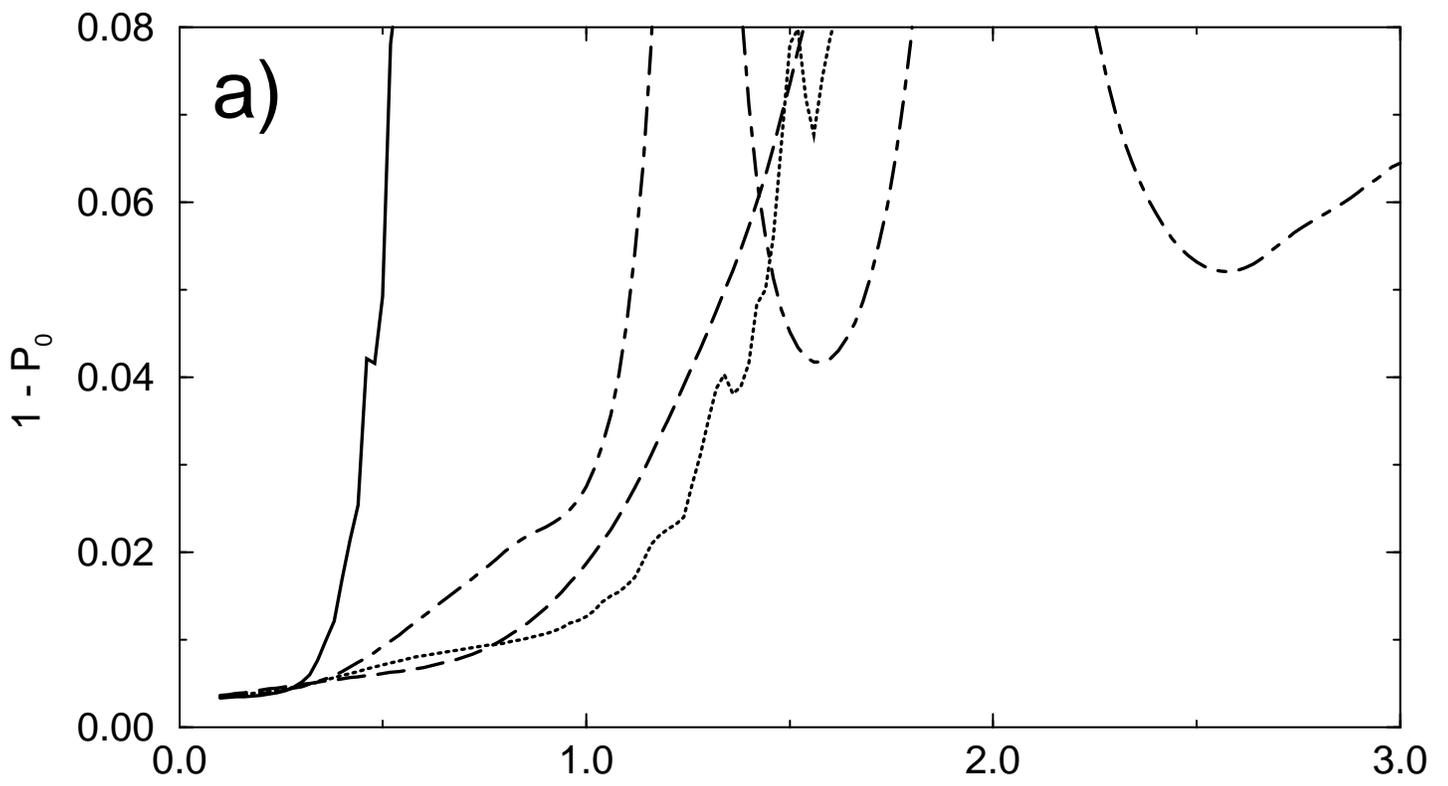
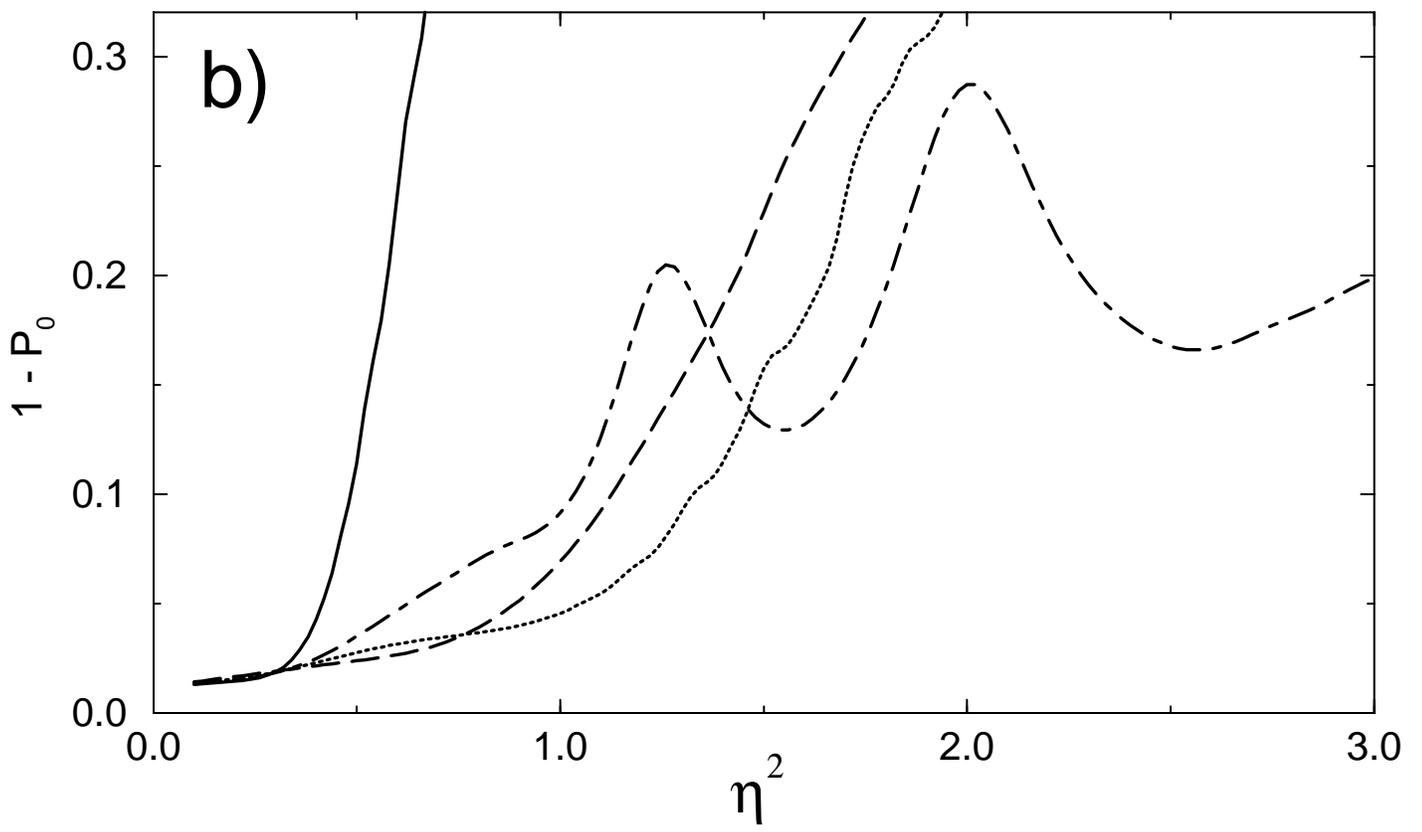
\end{document}